\documentclass[10pt,twocolumn,preprintnumbers,amsmath,amssymb,nofootinbib,superscriptaddress,prd]{revtex4-2}

\usepackage[dvipsnames]{xcolor}
\usepackage{bm}
\usepackage{graphicx}
\usepackage{tikz}
\usepackage[colorlinks,allcolors=RoyalBlue]{hyperref}
\usepackage[normalem]{ulem}
\newcommand{\eq}{\mathrm{eq}}
\newcommand{\gstar}{g_\star}
\newcommand{\gstars}{g_{\star S}}

\newcommand{\ma}{m_a} 
\newcommand{\fax}{f_a}

\newcommand{\mm}{{\rm MM}}
\newcommand{\nr}{{\rm GM}}
\newcommand{\fs}{{\rm FS}}
\newcommand{\iso}{{\rm iso}}
\newcommand{\mt}{{\rm MT}}

\newcommand{\Amm}{\alpha_\mm}
\newcommand{\Anr}{\alpha_\nr}
\newcommand{\Afs}{\alpha_\fs}
\newcommand{\Aiso}{\alpha_\iso}
\newcommand{\trap}{{\rm DO}}
\newcommand{\Atrap}{\alpha_\trap}
\newcommand{\coh}{{\rm DO}}

\newcommand{\pr}{{\rm PR}}
\newcommand{\Apr}{\alpha_\pr}

\begin{document}
\preprint{FERMILAB-PUB-25-0430-T}
\title{
Universal lower bound on the axion decay constant from free streaming effects
}

\author{Keisuke Harigaya}
\affiliation{Department of Physics, Enrico Fermi Institute, Leinweber Institute for Theoretical Physics, and Kavli Institute for Cosmological Physics, University of Chicago, Chicago, IL 60637, USA}
\affiliation{Kavli Institute for the Physics and Mathematics of the Universe (WPI), The University of Tokyo Institutes for Advanced Study, The University of Tokyo, Kashiwa, Chiba 277-8583, Japan}
\author{Wayne Hu}
\affiliation{Kavli Institute for Cosmological Physics, Enrico Fermi Institute, and Department of Astronomy \& Astrophysics, University of Chicago, Chicago IL 60637
}
\author{Rayne Liu}
\affiliation{Kavli Institute for Cosmological Physics, Enrico Fermi Institute, and Department of Astronomy \& Astrophysics, University of Chicago, Chicago IL 60637
}
\author{Huangyu Xiao}
\affiliation{Kavli Institute for Cosmological Physics, Enrico Fermi Institute, and Department of Astronomy \& Astrophysics, University of Chicago, Chicago IL 60637
}
\affiliation{Theory Division, Fermi National Accelerator Laboratory, Batavia, IL 60510, USA}

\date{\today}
\begin{abstract}
    We show that enhancement of the axion relic abundance compared to the standard misalignment contribution generically leads to the production of nonzero momentum axion modes, resulting in warm dark matter behavior and enhanced isocurvature perturbations.
    It leads to universal constraints on the axion parameter space that are independent of detailed model assumptions and cosmological history. For models enhancing relic abundance with gradient axion modes, observations of the Lyman-$\alpha$ forest impose a lower bound on the axion decay constant,  $\fax \gtrsim 10^{15} {\rm GeV}\,(10^{-18}{\rm eV}/\ma)$, from the free-streaming effect. For models relying on the delay of coherent axion oscillations, we obtain a slightly weaker bound, $\fax \gtrsim 10^{14} {\rm GeV}\,(10^{-18}{\rm eV}/\ma)$.
    We make relatively conservative choices to establish these universal bounds but also provide scaling parameters that can be calibrated for stronger constraints in concrete models and updated as observations improve. 
\end{abstract}
\maketitle

\section{Introduction}
Axions and axion-like particles (ALPs) are compelling dark matter candidates arising in many well-motivated extensions of the Standard Model, including the Peccei–Quinn solution to the strong CP problem \cite{Peccei:1977hh,PhysRevLett.40.223,PhysRevLett.40.279, Peccei:2006as,Abbott:1982af, Dine:1982ah,Preskill:1982cy} and string theory \cite{Svrcek:2006yi,Arvanitaki:2009fg}. 
The axion decay constant and mass are related in the QCD axion, predicting the so-called QCD axion line ($\ma-\fax$ relation). In the broader class of ALPs, the axion decay constant is independent of its mass, leaving more parameter space to be searched. If we require the axion density to reproduce the dark matter density, the viable ALP parameter space becomes a line in the standard vacuum misalignment mechanism. 

\begin{figure}
    \centering
    \includegraphics[width=1\linewidth]{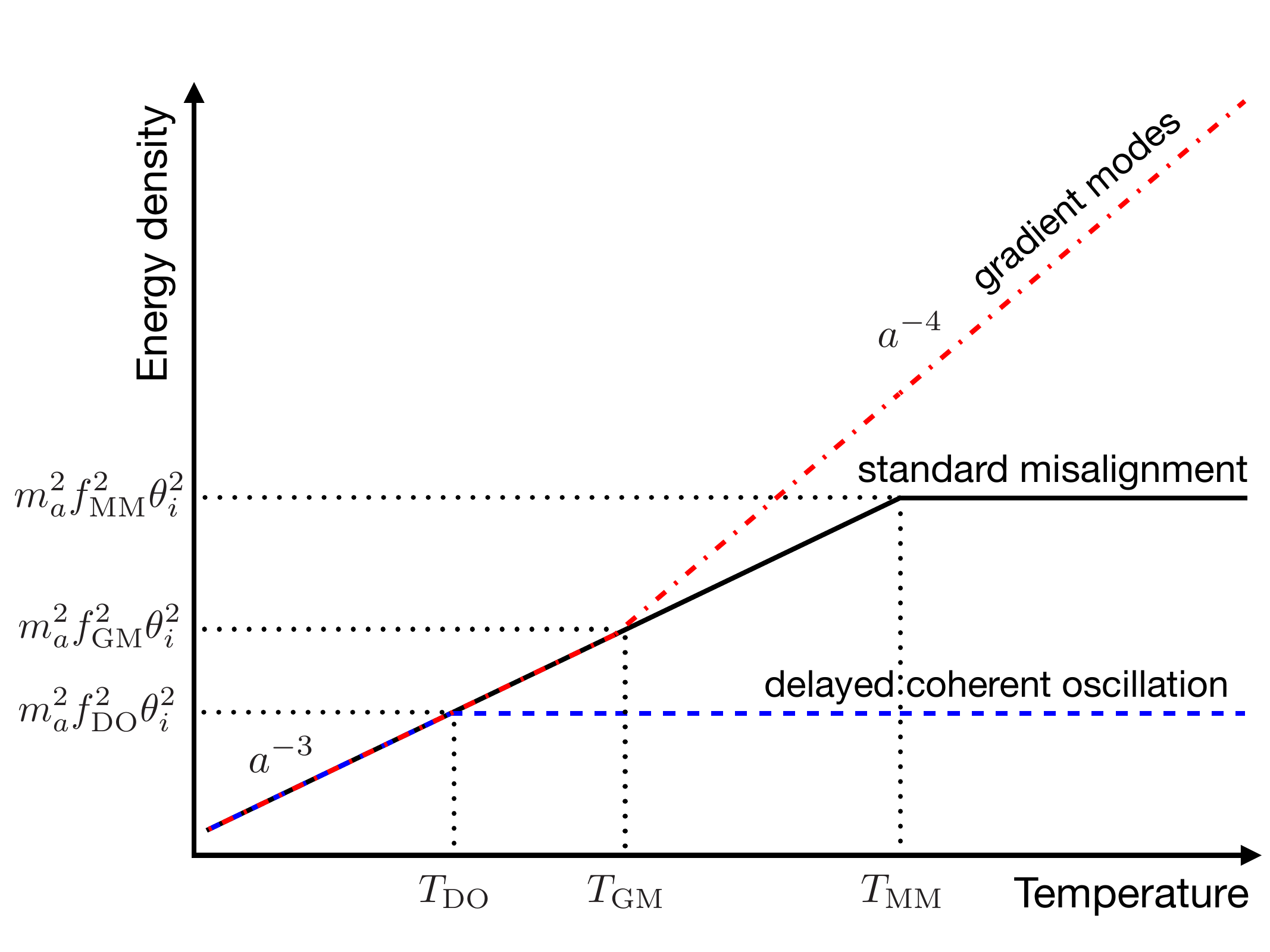}
    \caption{In the standard misalignment mechanism, the axion dark matter density originates from the potential energy and the axion begins oscillations when the axion mass $m_a \sim H$.
     Gradient modes (GM) or delayed beginning of coherent oscillations (DO) can enhance the axion abundance, allowing for a smaller decay constant $f_a$. 
    }    \label{fig:scaling}
\end{figure}

Current experimental searches or astrophysical probes of ALPs can only reach the axion parameter space with large couplings, which require an enhancement of their relic abundance compared to the standard misalignment mechanism to be the dark matter. There have been proposals to enhance the axion density,
which allows for a broader target for axion dark matter. Such enhancements typically involve delaying the onset of axion matter-like
oscillations through additional kinetic energy or gradient energy~\cite{Hiramatsu:2012sc,Co:2017mop,Baratella:2018pxi,Co:2019jts,Harigaya:2019qnl,Co:2020dya,Eroncel:2022vjg,Redi:2022llj,Harigaya:2022pjd,Eroncel:2024rpe,Eroncel:2025qlk,Bodas:2025eca} (dotted-dashed line in Fig.~\ref{fig:scaling}), leading to gradient modes, or through trapping of the axion field~\cite{Hook:2019hdk,DiLuzio:2021gos,Jeong:2022kdr,Papageorgiou:2022prc,DiLuzio:2024fyt} or taking the misalignment angle close to $\pi$~\cite{Turner:1985si,Strobl:1994wk,Co:2018mho,Takahashi:2019pqf,Arvanitaki:2019rax,Huang:2020etx} (dashed line in Fig.~\ref{fig:scaling}), which also produces gradient modes via parametric resonance as we will see.
These modes behave as warm or even relativistic dark matter at early times and can suppress the matter power spectrum through free-streaming, which can thus be constrained by observations.

There has been recent interest in the free-streaming effect of wave dark matter in general \cite{Amin:2022nlh,Liu:2024pjg,Ling:2024qfv,Long:2024imw,Liu:2025lts}, and when combined with isocurvature constraints leads to a model-independent mass bound on dark matter of post-inflation origin ($\ma\gtrsim10^{-19}$ eV)  from the Lyman-alpha forest \cite{Irsic:2017yje,Rogers:2020ltq}, which is 
weaker than the model-dependent ALP bound from isocurvature constraints alone \cite{Irsic:2019iff}, but 
stronger than the pre-inflationary fuzzy dark matter mass bound ($\ma>2\times 10^{-20}$ eV).

In this work, we will derive a universal bound on the axion decay constant as a function of the axion mass for a broad class of enhancement mechanisms.
Our study shows that a wide range of the axion parameter space, particularly for ultralight axions, is excluded by the requirement that axions behave as cold dark matter and remain consistent with large- and small-scale structure. These results apply broadly across axion models and provide a powerful, complementary constraint to direct searches and astrophysical probes.

\section{Vacuum Misalignment}
\label{sec:MM}

We first present the standard calculation on the axion zero-mode contribution to the relic abundance.
We consider the following Lagrangian of an axion $\phi$
\begin{equation}
    \mathcal{L} \supset\frac{1}{2}\partial_\mu \phi\partial^\mu \phi-m_a^2f_a^2\left(1-{\rm cos}\frac{\phi}{f_a}\right).
\end{equation}
It is convenient to define a dimensionless field value, $\theta = \phi/f_a$. In this work, the axion early-Universe dynamics and late-time observables are determined by the axion self-couplings from the cosine potential.
In Secs.~\ref{sec:MM}, \ref{sec:NR}, and\ \ref{sec:co}, we assume the axion mass $m_a$ is constant over cosmological time scale. A time-dependent axion mass is discussed in Sec.~\ref{sec:TD}.

In the vacuum misalignment mechanism, axions behave like matter after $\ma/H\sim 3$ and the amplitude starts to redshift away. Therefore, the relic abundance can be expressed as (e.g.~\cite{Turner:1985si})
\begin{equation}
    \Omega_a =  \Amm\frac{\ma^2\fax^2\theta_i^2}{2\rho_{\rm crit}}\frac{\gstars(T_0)T_0^3}{\gstars(T_\mm)T_{\mm}^3},
    \label{eq:Omegaa}
\end{equation}
where $T_0$ 
is the current CMB temperature, $\gstars$ is the effective entropic degrees of freedom, $T_\mm$ is defined as $3H(T_{\mm})= \ma$, and 
\begin{align}
\Amm={}&
\frac{8 (\Gamma[5/4])^2}{3\sqrt{3}\pi} 
\frac{1 + 0.523 x_\theta^{7/4}}{1+0.0588 x_\theta^{3/4}},\nonumber\\
x_\theta ={}&  \ln \frac{\pi}{\pi-|\theta_i|},
\label{eq:alphaMM}
\end{align}
accounts for the rolling of the axion field from an initial dimensionless value of $-\pi\le \theta_i\le \pi$ before oscillating around the minimum and includes a correction for large misalignment which has been fit to within $\sim 2\%$ of numerical results out to $x_\theta \sim 30$.  Note that $\lim_{x_\theta\rightarrow 0} \Amm \approx 0.4$.  

Given the dark matter abundance, we can invert Eq.~(\ref{eq:Omegaa}) to obtain the axion decay constant under the misalignment mechanism $f_\mm =\fax|_\mm$  as
\begin{align}
f_\mm ={}& 1.92\times 10^{16}{\, \rm GeV}
\frac{1}{|\theta_i|}  \left(\frac{10^{-18} \mathrm{eV}}{m_{a}}\right)^{1 / 4} \left(\frac{0.4}{\Amm}\right)^{1/2}
\nonumber\\
& \times \left[\frac{1}{100}\frac{\gstars^4 (T_{\mm})}{\gstar^3(T_{\mm})}\right]^{1/ 8}
 \left(\frac{ \Omega_a h^2}{0.12} \right)^{1/2},
 \label{eq:vacuum_misalignment}
\end{align}
where $\gstar$ is the effective degrees of freedom for the energy density.
Taking a large misalignment $\theta_i\rightarrow \pi$ or $\Amm \rightarrow \infty$ can in principle lower $f_\mm$ but note that the dependence is only logarithmic in the tuning of the initial field to the top of the potential.  For example even $\Amm=10$ or a reduction in $f_\mm$ by a factor of 5 requires 
$\pi/\theta_i-1=2.6\times 10^{-5}$.
Further more, large enhancement of the abundance is subject to the constraints discussed in Sec.~\ref{sec:co}.

Experiments often probe axion parameters at $\fax\ll f_{\mm}$ in Eq.~(\ref{eq:vacuum_misalignment}). While one could be agnostic about the production mechanism and assume that unknown dynamics in the early Universe will reproduce the correct relic abundance, physically  accounting for dark matter density with $\fax<f_{\mm}$ requires enhancing relic abundance.
In Secs.~\ref{sec:NR} and \ref{sec:co}, we discuss the enhancement by an extra gradient energy and a delay in the onset of axion matter, respectively.

\section{Gradient Axion Modes}
\label{sec:NR}

The axion abundance may be enhanced by extra gradient modes of axions.
Examples of such mechanism include 
parametric resonance~\cite{Co:2017mop,Harigaya:2019qnl,Co:2020dya},
the kinetic misalignment mechanism \cite{Co:2019jts,Eroncel:2022vjg},
 \footnote{In the kinetic misalignment mechanism, the axion field initially is coherent but fragments into non-zero modes unless $f_a$ is just below $f_{\mm}$~\cite{Eroncel:2022vjg}.}
the acoustic misalignment mechanism \cite{Bodas:2025eca,Eroncel:2025qlk}, long-lived domain walls \cite{Hiramatsu:2010yn,Hiramatsu:2012sc,Kawasaki:2014sqa,Baratella:2018pxi,Redi:2022llj,Harigaya:2022pjd}, etc.

\subsection{Axion abundance}

For gradient modes, let us assume that there is a characteristic comoving momentum of the axions $q_*$.   To infer the relic abundance, let us parameterize their number density  $n_a = \Anr m_a f_a^2/2$ at the scale factor $a_\nr = q_*/m_a$  when this gradient mode becomes nonrelativistic due to redshifting $q_* =a_\nr m_a$.   Then $f_\nr = \fax|_{q_*}$ is 
\begin{align}
f_\nr ={}&  1.24 \times 10^{15} {\, \rm GeV}\frac{1}{\Anr^{1/2}}
\left( \frac{10^{-18} \mathrm{eV}}{m_{a}}
\right)
\left(\frac {T_\nr}{10 \, \rm keV} \right)^{3/2}
\nonumber\\
{}&\times \left(\frac{ \gstars(T_\nr)}{4} \right)^{1/2} 
\left(\frac{ \Omega_a h^2}{0.12} \right)^{1/2}.
\label{eq:fnr}
\end{align}
Comparing Eq.~(\ref{eq:vacuum_misalignment}) and
(\ref{eq:fnr}) we see that $f_a$ can be reduced with a sufficiently low $T_\nr$, so that the axions are highly relativistic at $T_\mm \gg T_\nr$.
We shall see in the next section that in these scenarios bounds on free streaming and isocurvature modes constrain $T_\nr$ and hence $f_a$.

In general, the scaling parameter satisfies $\alpha_{\nr}\lesssim 1$ because axions need to behave like nonrelativistic matter at $T_\nr$ and hence the energy density of axions is at the most comparable to the largest possible potential energy density. Notably the acoustic misalignment mechanism can saturate $\alpha_\nr\approx 1$ \cite{Bodas:2025eca} while other models may lead to $\alpha_\nr\ll 1$. Now we would like to discuss the observational bounds on such gradient  modes required for enhancing the axion relic abundance.

\subsection{Free Streaming Bound}
\label{sec:fs}

Free streaming of dark matter particles or waves of a characteristic momentum $q_*$ suppresses structure below the maximal free streaming length.  Integrating the free streaming through to the matter dominated epoch, this scale is  \cite{Liu:2024pjg},
\begin{equation}
\lambda_\fs = \frac{\sqrt{2}}{a_\eq H_\eq} \frac{q_*}{a_\eq m_a} \ln\left( \frac{8 a_\eq m_a}{q_*} \right),
\end{equation}
in comoving coordinates.
As discussed above, we can reexpress this in terms of the scale factor at which the axions become nonrelativistic through $a_{\nr} = q_*/m_a$ to cast this in a mass independent way
\begin{equation}
\lambda_\fs = \frac{\sqrt{2}}{a_\eq H_\eq} \frac{a_\nr }{a_\eq} \ln\left( \frac{8 a_\eq }{a_\nr} \right).
\label{eq:lambdafs}
\end{equation}
A bound on the free streaming scale thus becomes a bound on how far before equality the dark matter must have been nonrelativistic.

For a monochromatic spectrum of a single momentum $q_*$, the suppression of the linear matter power spectrum due to free streaming can be characterized by a relative transfer function
\cite{Liu:2024pjg,Amin:2022nlh}
\begin{equation}
{\cal T}_{\rm rel}(k) = \frac{\sin(\lambda_\fs k)}{\lambda_\fs k}.
\end{equation}
This suppression is bounded by observations of small scale structure in the Universe, for example the Ly$\alpha$ forest in the translinear regime and bound structures in the fully nonlinear regime.  

The former provides conservative bounds on $a_\nr$ or equivalently $T_\nr$ that can be estimated directly from the transfer function.  The best studied case where free streaming is bounded by the Ly$\alpha$ forest is in the context of warm dark matter (WDM) where the relative transfer function integrated over the momentum distribution is given by~\cite{Viel:2005qj}
\begin{equation}
{\cal T}_{\rm WDM}(k) = [1+(\beta k)^{2\nu}]^{-5/\nu},
\end{equation}
where $\nu = 1.12$ and
\begin{equation}
\beta = 0.070 \left( \frac{\rm keV}{m_X} \right)^{1.11} 
\left( \frac{\Omega_X h^2}{0.12}\right)^{0.11}  {\rm Mpc}.
\end{equation}
Here $m_X$ is the mass of the WDM particle in the thermal scenario. We adopt the prescription from Ref.~\cite{Irsic:2017yje} for translating a bound on WDM to a case with a different shape to the relative transfer function by matching the scale $k_{3/4}$ at which the linear power reaches 3/4 of its unsuppressed value.   This prescription matches direct tests of WDM mass constraints and zero mode axion mass constraints.
We further take a bound on $m_X>3.1$\,keV from Ref.~\cite{Villasenor:2022aiy} who used $\Omega_X h^2\approx 0.12$.
This gives
\begin{equation}
\lambda_{\fs}< {0.084} {\rm Mpc}.
\end{equation} 
Note that if we instead matched the power at the $k_{1/2}$ point which is common in the literature, the bound would change negligibly to ${0.085}$\,{\rm Mpc} 
due to the similarity in the relative transfer function.
Using Eq.~(\ref{eq:lambdafs}), this puts a bound of 
\begin{equation}
\frac{a_\nr}{a_\eq} < 
5.07 \times 10^{-5}\left( \frac{\Omega_m h^2}{0.14} \right).
\end{equation}
To account for different spectra of momenta we can introduce a scaling parameter $\Afs$ and finally convert to temperature
\begin{equation}\label{eq:Tnr_fs}
T_\nr > 15.5 \Afs {\rm keV}.
\end{equation}
Following Ref.~\cite{Liu:2024pjg}, we consider a range of spectra defined by white noise field fluctuations with a constant power spectrum for $q<q_*$ and a $q^{-n}$  power law fall off for $q>q_*$.  For the range of power laws $n=4,\infty$, 
$0.8 < \Afs< 4$ approximately.  Note that the range extends to higher $\Afs$ more than lower since it only takes a small fraction of the dark matter  at higher momenta to impact the transfer function given the larger free streaming length.  Our fiducial choice of $\Afs=1$ is therefore on the conservative side.

Using Eq.~(\ref{eq:fnr}) this places a lower bound on $f_a$ from gradient-mode mechanisms as
\begin{align}
f_\nr >{}&  {2.36} \times 10^{15} {\, \rm GeV}\frac{\Afs^{3/2}}{\Anr^{1/2}}
\left( \frac{10^{-18} \mathrm{eV}}{m_{a}}
\right)
\left(\frac{ \Omega_a h^2}{0.12} \right)^{1/2},
\label{eq:fnr_fs}
\end{align}
where we have taken $\gstars(T_\nr)= \gstars(T_0)$ as appropriate for the evaluation of the lower bound.

\begin{figure}
    \centering
    \includegraphics[width=1\linewidth]{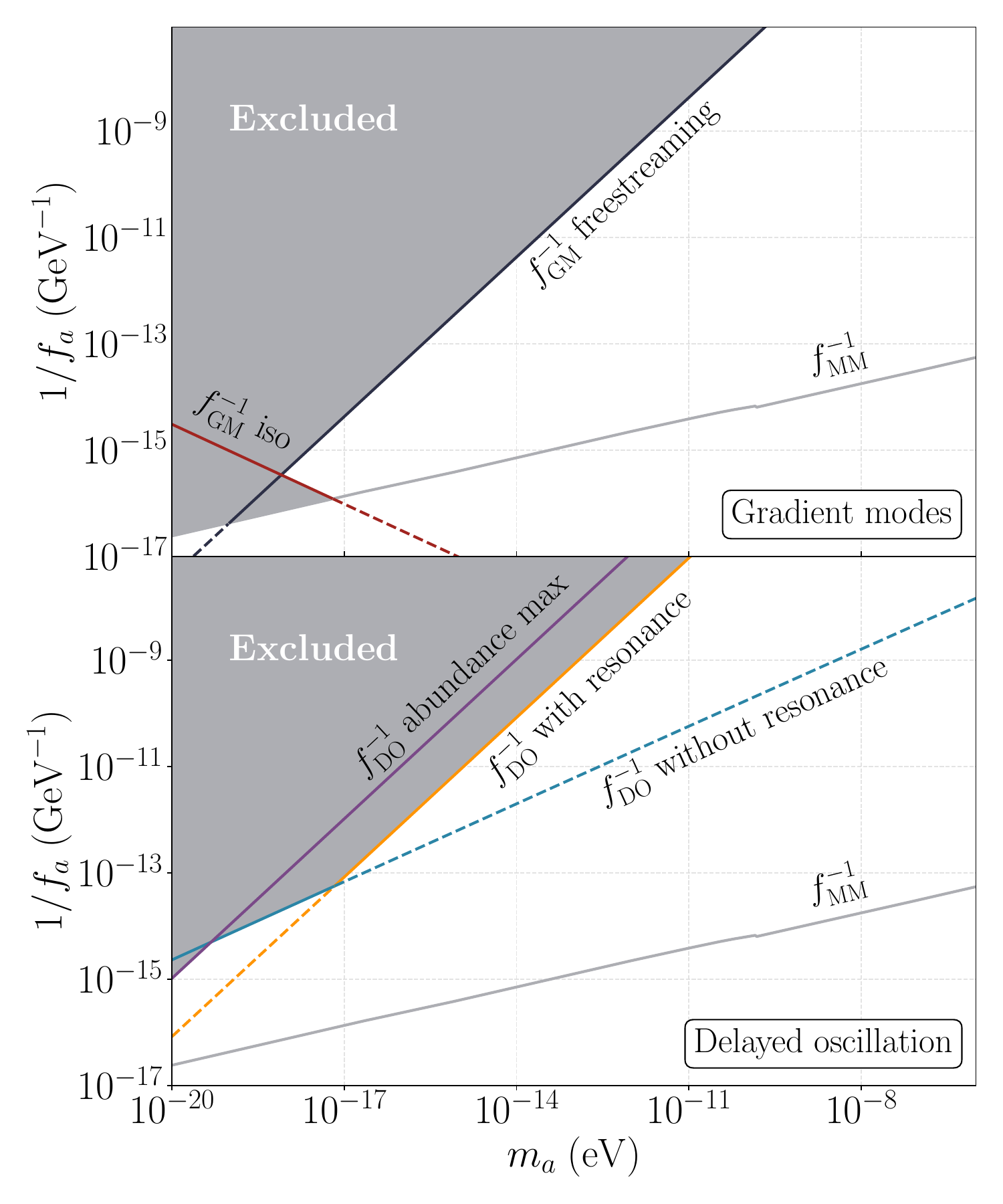}
    \caption{Constraints on $\fax^{-1}$ for models that raise it through gradient modes (top) and delayed coherent oscillations (bottom).  For $m_a \gtrsim 10^{-18}$\,eV free streaming bounds exclude $f_a^{-1} 
    \gtrsim 10^{-15}{\rm GeV}^{-1} (\ma/10^{-18}{\rm \,eV})$ and
    $\gtrsim 10^{-14}{\rm GeV}^{-1}(\ma/10^{-18}{\rm \,eV})$ respectively. 
    At lower masses the two mechanisms differ in their constraints.  For gradient modes, isocurvature bounds place a lower limit and for delayed oscillations, the efficiency of parameteric resonance and a limit on the maximal abundance from the minimal delay temperature $T_\coh$ provide the relevant bounds. 
    Dashed lines indicate the regions where one otherwise forbidden mechanism can be replaced by an allowed mechanism, e.g.~standard misalignment for $f_\mm^{-1}$.
    Our fiducial bounds employ relatively conservative choices of the 
    scaling parameters  $\Afs = 1, \Anr = 1, \Aiso = 1$, $\Amm(\theta_i=1)=0.43$, $\Atrap = 3, \Apr = 1$       (see text for discussion of ranges). }
        \label{fig:ourconstraints}
\end{figure}

\subsection{Isocurvature Bound}
If the dark matter is primarily composed of nonzero axion modes, the dark matter density is not initially nearly homogeneous on spatial scales $\sim 1/q_*$ as compared with the radiation.  This represents a dark matter isocurvature mode. 

The initial power spectrum of isocurvature perturbations will be peaked at $q_* = a_{\nr}m_a$ and constant or white for $k\ll q_*$ due to causality, which allows us to extrapolate the power spectrum to larger scales at $k_{\rm obs}\ll k_*$. Following Ref.~\cite{Irsic:2019iff}, we define the ratio of isocurvature to adiabatic fluctuations as 
\begin{equation}
    f^2_{\rm iso}
    = \frac{\Aiso}{A_s}\left(\frac{k_{\rm CMB}}{a_\nr m_a}\right)^3,
\end{equation}
where  $k^3 P_{\zeta}/2\pi^2 = A_s (k/k_{\rm CMB})^{n_s-1}$ with an amplitude  $A_s=2.054\times 10^{-9}$ at 
$k_{\rm CMB}= 0.05\rm\,Mpc^{-1}$.
 $\Aiso$  is the analogous normalization of the white noise power extrapolated to $q_*$, 
 $k^3 P_{\rm iso}/2\pi^2 = (k/q_*)^3 \Aiso$. Note that the actual isocurvature power spectrum need not be white out to $k\sim q_*$ as long as it is white between $k_{\rm CMB}$ and the observational scale of the bound.
 Relatively less power at high $k$ is possible in many scenarios \cite{Bodas:2025eca,OHare:2021zrq},
and given rms density $\delta_{\rm rms}$, taking $\Aiso\sim \delta_{\rm rms}^2$ can then substantially underestimate the actual amplitude at $k_{\rm CMB}$. 
Conversely, $\delta_{\rm rms}$ itself may not reach unity for all mechanisms.
  Therefore, we adopt $\Aiso= 1$ for our benchmark limits but take $0.1-100$ as a plausible range.  Current Ly$\alpha$ constraints require $f_{\rm iso}<3\times 10^{-3}$ which equates to 
\begin{equation}
 \frac{a_{\rm NR}}{a_{\rm eq}}>4.07 \times 10^{-5}\Aiso^{1/3}\left(\frac{10^{-18}\,\rm eV}{m_a}\right)\left( \frac{\Omega_m h^2}{0.14}\right),
\end{equation}
or
\begin{equation}\label{eq:Tnr_iso}
    T_{\nr}< 19.3 {\,\rm keV}\, \Aiso^{-1/3}
    \left(\frac{m_a}{10^{-18}\rm\,eV}\right),
\end{equation}
where we have used $\gstars(T_{\nr})= \gstars(T_0)$ given the low keV temperatures involved. 
Unlike the free streaming bound, this places an upper limit on $T_{\rm NR}$ and the combined bound gives
\begin{equation}
m_a 
> 8.0 \times 10^{-19}{\, \rm eV} \, \Afs \Aiso^{1/3}.
\end{equation}
Note that this mass bound does not specifically require the dark matter to be an axion but does require the large stochastic fluctuations in density that occurs in the gradient mode axion mechanism (see \cite{Amin:2022nlh}).\footnote{We take the free streaming and isocurvature bounds as independent constraints under the assumption that where the isocurvature bound is stronger, the free streaming scale is below the Ly$\alpha$ scale so that isocurvature modes grow gravitationally in matter domination \cite{Liu:2025lts,Amin:2025sla} and when the free streaming bound is stronger, isocurvature modes can be ignored.  When they are comparable the competition between suppressed growth of the additional isocurvature modes and free streaming reduction of the adiabatic modes can change the shape of the matter power spectrum and the details of the joint constraint.}
When applied to the axion, the isocurvature bound through Eq.~(\ref{eq:fnr}) implies an upper bound on $f_a$ that is higher than that of $f_\mm$ for masses much larger than this minimum 
\begin{equation}
f_\nr < 3.28\times 10^{15}{\,\rm GeV} \,
\left(\frac{1}{\Aiso \Afs} 
\frac{m_a}{10^{-18}{\,\rm eV}}
\frac{ \Omega_a h^2}{0.12} \right)^{1/2}.
\end{equation}
Once this bound crosses $f_\mm$ then the allowed models in the $\fax-m_a$ plane that remain are zero mode models, e.g., for PQ symmetry breaking before inflation.
In Fig.~\ref{fig:ourconstraints} (upper panel), we plot the axion parameter space constrained by the free-streaming effect and enhanced isocurvature perturbations, and the shaded region is excluded with relatively conservative choices of $\alpha$ scaling parameters: $\Afs = 1, \Anr = 1, \Aiso = 1$ with $\Amm(\theta_i=1)=0.43$ computed from Eq.~(\ref{eq:alphaMM}).
Values for $\gstars (T_\mm)$ and $\gstar(T_\mm)$ are taken from \cite{Husdal:2016haj}, Tab.\ A1.
The region below $f_\mm^{-1}$ curve is not excluded since zero modes can explain the relic density for which the free streaming and isocurvature bounds do not apply.

The ranges of scaling parameters discussed above affect the bounds as follows:   for $0.8 < \Afs < 4$ the upper bound on $f_a^{-1}$ varies, mostly in the stronger direction, by a factor of
$0.7-8$; for $\Anr\ll 1$ the bound strengthens as $\Anr^{1/2}$;
for $0.1<\Aiso<100$ the lower bound on $f_a^{-1}$ varies by $0.3-10$, again mostly in the stronger direction. 
A fine tuning of initial misalignment angle to $10^{-5}$ similarly would only increase $f_\mm^{-1}$ by a factor of 5 which would only affect the allowed region below $m_a=10^{-18}$\,eV. 

In summary, the excluded region in Fig.~\ref{fig:ourconstraints} (upper panel) gives a  model-independent bound on axion dark matter whose relic density is enhanced through introducing gradient energy that errs on the conservative side.

\section{Delayed coherent oscillations}
\label{sec:co}

Another way to enhance the axion abundance in comparison with the misalignment production is to delay the beginning of the coherent oscillation by taking $|\theta_i-\pi| \ll 1$ or trapping the axion field at $\theta\neq 0$ and releasing it at $T = T_{\trap} \ll T_\mm$. We have already seen in Eq.~(\ref{eq:alphaMM}) that a large initial misalignment which retains fully coherent oscillations only gives a logarithmic enhancement as $\theta_i\rightarrow \pi$. For the trapping case, possible UV completions can be found in~\cite{Hook:2019hdk,DiLuzio:2021gos,Jeong:2022kdr,Papageorgiou:2022prc,DiLuzio:2024fyt}.

\subsection{Axion abundance}
To produce the correct relic abundance, the axion decay constant is required to be
\begin{align}
\label{eq:fcoh}
f_\coh ={}&  
1.24 \times 10^{15} {\, \rm GeV}\frac{1}{|\theta_i|}\left( \frac{10^{-18} \mathrm{eV}}{m_{a}}\right)\\
{}&\times \left(\frac {T_\trap}{10 \, \rm keV} \right)^{3/2} \left(\frac{ \gstars(T_\trap)}{4} \right)^{1/2} \left(\frac{ \Omega_a h^2}{0.12} \right)^{1/2}.\nonumber
\end{align}
$T_\trap$ cannot be arbitrarily small.
Since dark matter does not exist at $T>T_\trap$, we expect $O(1)$ differences in the matter power spectrum for  length scales shorter than the horizon size at $T= T_{\trap}$, e.g.~\cite{Niedermann:2020dwg}.
We generically expect that the Ly$\alpha$ forest constraint on the 3/4 power scale $k_{3/4}$ from \S \ref{sec:fs} provides a constraint 
 \begin{equation}
k_{3/4} >  {a_\trap H(T_{\trap})}  \Atrap ,
 \end{equation}
where $\Atrap={\cal{O}}(1)$ depends on the specific details of the trapping mechanism,
which we again infer from the WDM bound to be
\begin{equation}
 T_\trap> 1.19 {\rm keV} /\Atrap.
 \label{eq:Ttrapbound}
\end{equation}
We take $\Atrap=3$ as a fiducial value but consider a very conservative range $1 \le \Atrap \le 100$.

Using $\theta_i <1$ for the maximum abundance and the bound in Eq.~\eqref{eq:Ttrapbound}, we obtain
\begin{align}
\label{eq:fcoh3}
    f_\coh > 5.0\times 10^{13} 
    {\, \rm GeV}\frac{1}{\Atrap^{3/2}}\left( \frac{10^{-18} \mathrm{eV}} {m_{a}} \right)\left(\frac{ \Omega_a h^2}{0.12} \right)^{1/2},
\end{align}
which must be satisfied on top of all the other criteria discussed below.  Here and below  $\gstar$ and $\gstars$ are taken at their low-temperature limits.

\subsection{Free Streaming Bound}

Although the axion can be initially coherent, the self-interaction of the axion can produce axions in non-zero modes.
The linearized equation of motion of the fluctuations $\phi_q$ right after the release is
\begin{align}
    \ddot{\phi_q} + \left(\frac{q^2}{a_\trap^2} + m_a^2 - \frac{1}{2} \theta_i^2 m_a^2 {\cos}^2[m_a (t-t_\trap) ]\right)\phi_q=0.
\end{align}
Here we neglect the cosmic expansion on the oscillation time scale, 
take the small-amplitude limit,
and hence 
take the zero mode
oscillation of the axion as sinusoidal.
Recasting the equation of motion to Mathieu's equation, we  find that the parametric resonance occurs with a rate of $\Gamma \simeq \frac{1}{16} \theta_i^2 m_a$ and produces axions with a comoving wave number
\begin{align}
   q_* \simeq \frac{\Apr}{2} \theta_i m_a a_\trap,
\end{align}
where $\Apr= {\cal O}(1)$ for the resonance band and potentially larger if the resonantly produced axions efficiently upscatter.

The condition for producing enough nonzero modes through parametric resonance that have comparable energy to zero-modes is
\begin{align}
    \frac{\Gamma}{H(T_\trap)}
    =\frac{3\sqrt{10}}{16 \pi\gstar^{1/2}}
   \frac{ m_a M_{\rm pl} }{T_\trap^2} \theta_i^2
    > {}&  {\rm ln}\frac{f_a}{m_a\theta_i^{1/2}},
    \label{eq:resonancecondition}
\end{align}
where we conservatively assume that the initial seeds of the fluctuations are given by quantum fluctuations to get the right hand side.  Note that given the large hierarchy between $\fax$ and $m_a$, it is a good approximation to drop $\theta_i$ in logarithm as well as $O(1)$ coefficients implicit in this approximation.
When the above condition is satisfied, we will have a free streaming bound due to the finite $q_*$ momenta.

Let us first consider the case where the resonance condition (\ref{eq:resonancecondition}) is satisfied. 
With the assumption that the resonant axions at $q_* \approx a_\trap \Apr \theta_i m_a/2$ are nonrelativistic at $a_\trap$,\footnote{Relativistic axions would only strengthen the free streaming bound as well as require a larger $\fax$ to compensate for the kinetic energy lost to redshifting.} 
we can estimate the free streaming bound by integrating a non-relativistic velocity $v= q_*/am_a$ from $a_\trap$ to give 
\begin{align}
\lambda_\fs ={}& \frac{1}{\sqrt{2} a_\eq H_\eq} \frac{a_\trap \Apr \theta_i }{a_\eq} \ln\left( \frac{4 a_\eq }{a_\trap} \right)\nonumber\\
<{}& 0.084 {\rm Mpc}.
\label{eq:lambdatrap}
\end{align}
Up to the log term, the trapping constraint behaves similarly to the gradient constraint but with $a_\nr \rightarrow a_\trap\alpha_\pr \theta_i/2 \propto \theta_i/T_\trap$.
Notice that the free streaming scale  is suppressed for small
$\theta_i/T_\trap$ and we can place a  lower bound on $f_\coh$ from the lower bound on $T_\trap$ in Eq.~(\ref{eq:Ttrapbound}). 

Now let us consider the case where resonance does not occur.  The relic abundance is still given by Eq.~(\ref{eq:fcoh}) but the axions are in the zero mode.   The condition for resonance not occurring in Eq.~(\ref{eq:resonancecondition}) also 
requires small $\theta_i/T_\trap$ and therefore the lower bound on $T_\trap$ in Eq.~(\ref{eq:Ttrapbound}) again puts a lower bound on $f_\coh$.

With these bounds, models are allowed either if they satisfy the free streaming bound with resonance 
\begin{align}
\label{eq:fcoh1}
f_\coh > {}&
2.73 \times 10^{13} {\, \rm GeV}\frac{\Apr}{\Atrap^{1/2} }\left( \frac{10^{-18} \mathrm{eV}}{m_{a}}\right)\\
{}&\times \ln\left(\frac{6036}{\Atrap}\right) \left(\frac{ \Omega_a h^2}{0.12} \right)^{1/2},\nonumber
\end{align}
or if resonance does not occur
\begin{align}
\label{eq:fcoh2}
f_\coh >{}& 6.68 \times 10^{14} {\, \rm GeV} \frac{1}{\Atrap^{1/2} }\left( \frac{10^{-18} \mathrm{eV}}{m_{a}}\right)^{1/2} \nonumber\\
{}&\times
 \left[ \ln \left( \frac{\fax}{m_a} \right) \right]^{-1/2}\left(\frac{ \Omega_a h^2}{0.12} \right)^{1/2},
\end{align}
where the implicitness of the log term can be iteratively resolved.  
If either of the above condition is satisfied, it is possible to arrange a $\theta_i$ and $T_\trap$ to make the parameter space viable.

In Fig.~\ref{fig:ourconstraints} (lower panel), we show the resultant exclusion region for $\Atrap=3$ and $\Apr=1$.  For $m_a\lesssim 10^{-18}$\,eV the lack of parameteric resonance sets the more relevant bound and for $m_a\lesssim 10^{-19}$ even without resonance, the axion abundance itself from 
Eq.~(\ref{eq:fcoh3})
is insufficient given the lower limit on $T_\trap$ from Eq.~(\ref{eq:Ttrapbound}).  For the range $1\le \Atrap \le 100$ 
the bound on $\fax^{-1}$ changes by a factor of $0.5-10$ with resonance and $0.6-6$, where the higher values are very conservative.  Parameteric resonance and scattering that produces $\Apr>1$ would strengthen the resonance upper bound as $\Apr^{-1}$.

Finally,
in deriving the bounds in Eq.~\eqref{eq:fcoh1} and \eqref{eq:fcoh2}, $\theta_i$ is taken to be the most optimal value to evade the free-streaming bound to derive a model-independent conservative bound. 
In a specific class of models, however, $\theta_i$ is not a free parameter but is a prediction of the theory~\cite{DiLuzio:2021gos}. More generally, small $\theta_i$ may require fine-tuning of the parameters of theories. For a fixed $\theta_i$, the lower bounds from free streaming are given approximately by
\begin{equation}
T_\coh > 0.726{\, \rm keV}
\Apr \theta_i \ln(10885 \Apr \theta_i)
\end{equation}
provided $T_\coh$ is higher than the bound in Eq.~(\ref{eq:Ttrapbound}).  The coefficients here are set to match the mathematical result at $\Apr\theta_i=0.1,1$ and the form is a good approximation at the few percent level down to $\Apr\theta_i \sim 10^{-2}$, whereas $\Apr\theta_i>1$ is physically limited by the use of the non-relativistic approximation to the resonant axions.
Using this bound in Eq.~(\ref{eq:fcoh}), we obtain 
\begin{align}
    f_\coh > 
{}&  
2.40 \times 10^{13} {\, \rm GeV}|\theta_i|^{1/2}\Apr^{3/2}\left( \frac{10^{-18} \mathrm{eV}}{m_{a}}\right)\\
{}&\times 
\left[ \ln\left( 10885 \Apr\theta_i\right)\right]^{3/2} \left(\frac{ \Omega_a h^2}{0.12} \right)^{1/2}\nonumber
\end{align}
to satisfy the free streaming bound with resonance.  Likewise models that avoid resonance exist if instead  
\begin{align}
T_\coh > 15.8 {\, \rm keV}
\theta_i \left( \frac{m_a}{10^{-18} \mathrm{eV}}\right)^{1/2} 
\left[\ln \left(\frac{\fax}{m_a}\right)\right]^{-1/2}
\end{align}
and
\begin{align}
    f_\coh > 
{}&  
2.44 \times 10^{15} {\, \rm GeV}\frac{1}{|\theta_i|}\left( \frac{10^{-18} \mathrm{eV}}{m_{a}}\right)^{1/4}\\
{}&\times 
\left[\ln\left(\frac{\fax}{m_a} \right)\right]^{-3/4} \left(\frac{ \Omega_a h^2}{0.12} \right)^{1/2}\nonumber
\end{align}
provided they satisfy Eq.~(\ref{eq:Ttrapbound}).

\begin{figure}
    \centering
    \includegraphics[width=1\linewidth]{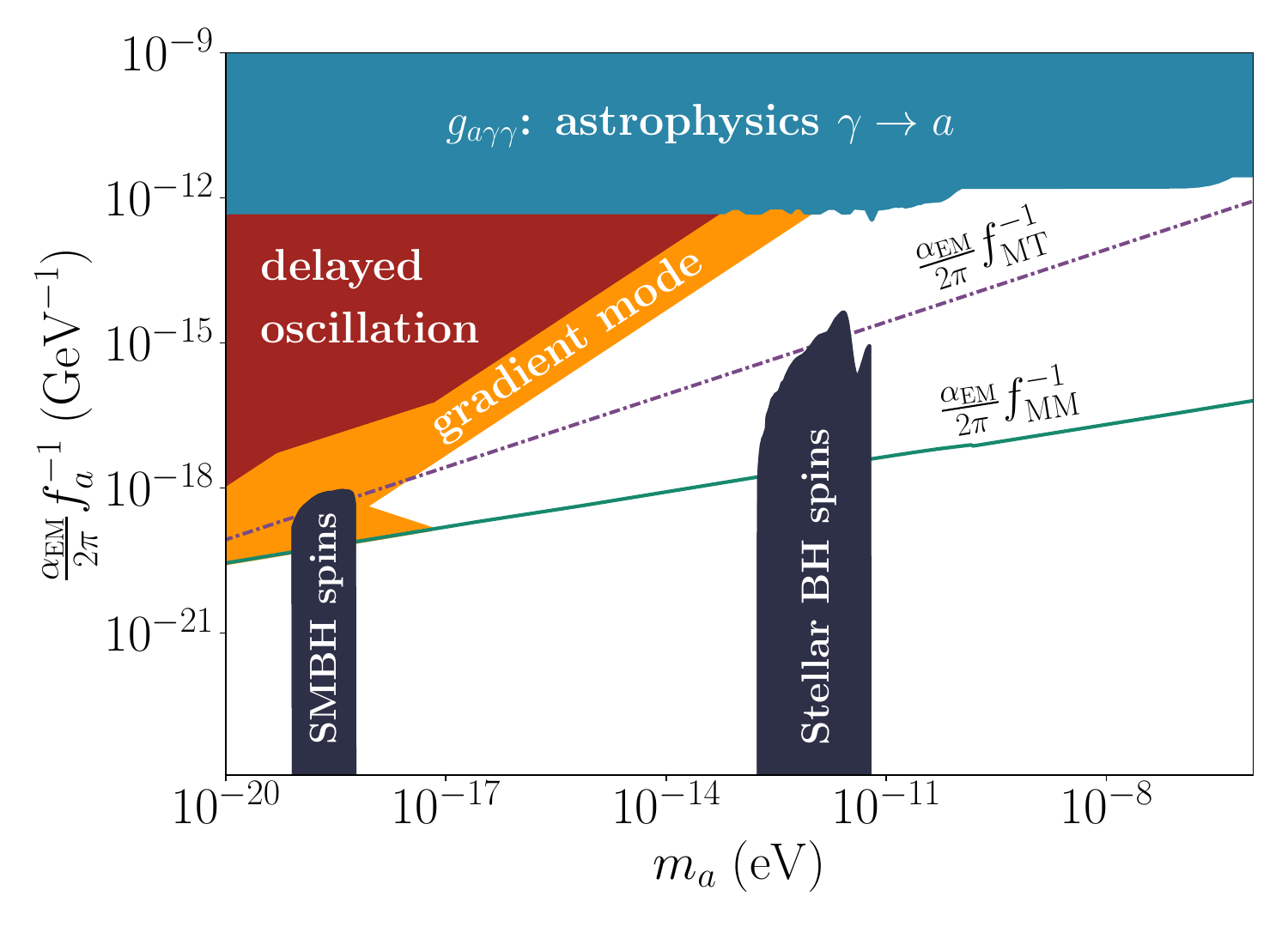}
    \caption{Bounds on gradient modes and delayed oscillations from Fig.~\ref{fig:ourconstraints} combined with those on the axion-photon coupling assuming  $g_{a\gamma\gamma}=\alpha_{\rm EM}/(2\pi f_a)$ from astrophysical photon conversion $\gamma\rightarrow a$ and blackhole (BH) superradiance compiled in \cite{AxionLimits}, neither of which require the axions to be the dark matter.   The excluded region for $m_a \lesssim 10^{-18}$\, eV and $f_a^{-1}< f_\mt^{-1}$ are viable only if the mass is sufficiently temperature dependent to enhance the axion abundance from the misalignment mechanism, $f_\mm^{-1}$  (see Eq.~(\ref{eq:fmt})).}
    \label{fig:constraints_gagg}
\end{figure}

\section{Time-dependent mass}
\label{sec:TD}

We have so far considered an axion mass that is constant over cosmological time.
We discuss the case where the axion mass adiabatically increases and show that the lower bound on $f_a$ 
cannot essentially
be relaxed beyond the limits set in Secs.~\ref{sec:NR} and \ref{sec:co}.%
\footnote{
The mass may change non-adiabatically, which is possible, for example, if the strong dynamics responsible for the axion mass exhibits a first-order phase transition. Constraints similar to the delayed coherent oscillation case are expected to be applicable with $T_\coh$ interpreted as the temperature at which the mass changes non-adiabatically, since the perturbations produced before the non-adiabatic change are modified from the $\Lambda$CDM prediction.
The dynamics before the non-adiabatic change may put stronger constraints, which can be investigated model dependently.
}
The adiabatic increase of an axion mass is possible if the axion mass is generated by strong dynamics and the temperature of the bath of the strongly coupled sector is above the dynamical scale.

In the zero mode case of Sec.~\ref{sec:MM}, if the axion mass changes after the beginning of oscillations at $m_a(T_\mt)=3H(T_\mt)$ before reaching its final value of $m_a$, the axion abundance is modified and the modification can be expressed by~\cite{Lyth:1991ub}
\begin{align}
\label{eq:enhance_MT}
    \Amm \rightarrow \Amm \sqrt{\frac{ m_a}{3H(T_\mt)}},
\end{align}
with minor changes to the scaling of $\Amm$ with $\theta_i$ in Eq.~(\ref{eq:alphaMM}).
We then obtain the $m_a(T)$ generalization of $f_\mm$,
\begin{align}
\label{eq:fmt}
f_\mt ={}&  4.84\times 10^{15}{\, \rm GeV}
\frac{1}{|\theta_i|}  \left(\frac{10^{-18} \mathrm{eV}}{m_{a}} \frac{T_\mt}{\rm keV}\right)^{1 / 2} 
\nonumber\\
& \times
\left(\frac{0.4}{\Amm} \frac{ \Omega_a h^2}{0.12} \right)^{1/2}
\left[\frac{1}{100}\frac{\gstars^2 (T_{\mt})}{\gstar(T_{\mt})}\right]^{1/ 4}.
\end{align}

The lower bound on $T_{\mt}$ should be similar to the lower bound on $T_{\trap}$ which raises $f_\mt^{-1}$ 
but not enough to change the constraints in Fig.~\ref{fig:ourconstraints}  above $m_a \gtrsim 10^{-18}$eV (see Fig.~\ref{fig:constraints_gagg}).

In the gradient modes case of Sec.~\ref{sec:NR},
if the axion mass increases after the axion fluctuations become non-relativistic, the axion abundance is suppressed by $m_a(T_{\nr})/ m_a(T=0)$ in comparison with the case where the axion mass is constant. 
Therefore, the lower bound on $f_a$ becomes stronger.%
\footnote{
If the mass decreases adiabatically, the gradient modes and delayed coherent oscillation case may evade the bounds, providing a possible loophole. Such scenarios will be still subject to free-streaming and isocurvature constraints, which can be derived model-dependently. The bound for the zero mode case become stronger.
}

In the delayed coherent oscillation case of Sec.~\ref{sec:co}, 
the axion abundance is also suppressed by $m_a(T_{\trap})/ m_a(T=0)$  and the lower bounds on $f_{\coh}$ in Eqs.~\eqref{eq:fcoh3} and \eqref{eq:fcoh1} become stronger.
One can show that the condition for the parametric resonance to occur is independent of $m_a(T_{\trap})$ after choosing $\theta_i$ to explain the observed dark matter density, so Eq.~\eqref{eq:fcoh2} is not altered.

\section{Discussion}

In this paper, we derived a universal lower bound on the decay constant of dark-matter axion. We showed that significant enhancement of an axion dark-matter abundance in comparison with the misalignment contribution generically leads to production of non-zero modes of axions and is constrained by the free-streaming effect.
For gradient modes, this leads to a bound of
 $\fax \gtrsim 10^{15} {\rm GeV}\,(10^{-18}{\rm eV}/\ma)$ and for the parametric resonance modes $\fax \gtrsim 10^{{14}} {\rm GeV}\,(10^{-18}{\rm eV}/\ma)$ for relatively conservative choices of model dependent parameters which we have absorbed into a series of scaling parameters.  Note that the scaling parameters $\Afs$, $\Aiso$ and $\Atrap$ can also be used in the future to scale constraints to improved observations of the Ly$\alpha$ forest, isocurvature modes, and minimum temperature of axion production respectively.

The lower bound can be interpreted as an upper bound on the axion coupling to standard model particles. Assuming that the shift symmetry of the axion field has electromagnetic anomaly, the axion couples to photons,
\begin{align}
{\cal L} \supset \frac{g_{a \gamma \gamma}}{4}a F\tilde{F},~~g_{a\gamma \gamma} \sim \frac{\alpha}{2\pi f_a}. 
\end{align}
The exclusion regions from Fig.~\ref{fig:ourconstraints} on gradient modes and delayed coherent oscillations are translated to 
$g_{a \gamma \gamma}$ in Fig.~\ref{fig:constraints_gagg}.
The excluded region below the $f_{\mt}^{-1}$ line and below $m_a\lesssim 10^{-18}$eV  become viable only if the mass varies with temperature sufficiently (see Eq.~\eqref{eq:fmt}).
Our bound is complementary to other  bounds, especially these astrophysical bounds from photon-axion conversion and black hole superradiance that do not require the axion to be the dark matter.

Our model-independent discussion will be also useful in constraining concrete models, where parameters such as $T_{\nr}$, $T_{\coh}$, and $\theta_i$ may be subject to stronger model-dependent constraints. The scaling factors we have introduced to characterize the physical quantities involved in the bound  can be precisely determined once the detail of the model is fixed.

\acknowledgments

  We thank Austin Joyce for useful comments.
K.H. is supported by World Premier International Research Center Initiative (WPI), MEXT, Japan (Kavli IPMU).
W.H.\ \& R.L. are supported by U.S.\ Dept.\ of Energy contract DE-FG02-13ER41958 and the Simons Foundation. H.X.~is supported by Fermi Forward Discovery Group, LLC under Contract No.\ 89243024CSC000002 with the U.S.\ Dept.\ of Energy, Office of Science, Office of High Energy Physics.

\eject
\onecolumngrid
\appendix
\bibliographystyle{apsrev4-2}
\bibliography{main.bbl}

\begin{thebibliography}{50}%
\makeatletter
\providecommand \@ifxundefined [1]{%
 \@ifx{#1\undefined}
}%
\providecommand \@ifnum [1]{%
 \ifnum #1\expandafter \@firstoftwo
 \else \expandafter \@secondoftwo
 \fi
}%
\providecommand \@ifx [1]{%
 \ifx #1\expandafter \@firstoftwo
 \else \expandafter \@secondoftwo
 \fi
}%
\providecommand \natexlab [1]{#1}%
\providecommand \enquote  [1]{``#1''}%
\providecommand \bibnamefont  [1]{#1}%
\providecommand \bibfnamefont [1]{#1}%
\providecommand \citenamefont [1]{#1}%
\providecommand \href@noop [0]{\@secondoftwo}%
\providecommand \href [0]{\begingroup \@sanitize@url \@href}%
\providecommand \@href[1]{\@@startlink{#1}\@@href}%
\providecommand \@@href[1]{\endgroup#1\@@endlink}%
\providecommand \@sanitize@url [0]{\catcode `\\12\catcode `\$12\catcode `\&12\catcode `\#12\catcode `\^12\catcode `\_12\catcode `\%12\relax}%
\providecommand \@@startlink[1]{}%
\providecommand \@@endlink[0]{}%
\providecommand \url  [0]{\begingroup\@sanitize@url \@url }%
\providecommand \@url [1]{\endgroup\@href {#1}{\urlprefix }}%
\providecommand \urlprefix  [0]{URL }%
\providecommand \Eprint [0]{\href }%
\providecommand \doibase [0]{https://doi.org/}%
\providecommand \selectlanguage [0]{\@gobble}%
\providecommand \bibinfo  [0]{\@secondoftwo}%
\providecommand \bibfield  [0]{\@secondoftwo}%
\providecommand \translation [1]{[#1]}%
\providecommand \BibitemOpen [0]{}%
\providecommand \bibitemStop [0]{}%
\providecommand \bibitemNoStop [0]{.\EOS\space}%
\providecommand \EOS [0]{\spacefactor3000\relax}%
\providecommand \BibitemShut  [1]{\csname bibitem#1\endcsname}%
\let\auto@bib@innerbib\@empty
\bibitem [{\citenamefont {Peccei}\ and\ \citenamefont {Quinn}(1977)}]{Peccei:1977hh}%
  \BibitemOpen
  \bibfield  {author} {\bibinfo {author} {\bibfnamefont {R.~D.}\ \bibnamefont {Peccei}}\ and\ \bibinfo {author} {\bibfnamefont {H.~R.}\ \bibnamefont {Quinn}},\ }\href {https://doi.org/10.1103/PhysRevLett.38.1440} {\bibfield  {journal} {\bibinfo  {journal} {Phys. Rev. Lett.}\ }\textbf {\bibinfo {volume} {38}},\ \bibinfo {pages} {1440} (\bibinfo {year} {1977})}\BibitemShut {NoStop}%
\bibitem [{\citenamefont {Weinberg}(1978)}]{PhysRevLett.40.223}%
  \BibitemOpen
  \bibfield  {author} {\bibinfo {author} {\bibfnamefont {S.}~\bibnamefont {Weinberg}},\ }\href {https://doi.org/10.1103/PhysRevLett.40.223} {\bibfield  {journal} {\bibinfo  {journal} {Phys. Rev. Lett.}\ }\textbf {\bibinfo {volume} {40}},\ \bibinfo {pages} {223} (\bibinfo {year} {1978})}\BibitemShut {NoStop}%
\bibitem [{\citenamefont {Wilczek}(1978)}]{PhysRevLett.40.279}%
  \BibitemOpen
  \bibfield  {author} {\bibinfo {author} {\bibfnamefont {F.}~\bibnamefont {Wilczek}},\ }\href {https://doi.org/10.1103/PhysRevLett.40.279} {\bibfield  {journal} {\bibinfo  {journal} {Phys. Rev. Lett.}\ }\textbf {\bibinfo {volume} {40}},\ \bibinfo {pages} {279} (\bibinfo {year} {1978})}\BibitemShut {NoStop}%
\bibitem [{\citenamefont {Peccei}(2008)}]{Peccei:2006as}%
  \BibitemOpen
  \bibfield  {author} {\bibinfo {author} {\bibfnamefont {R.~D.}\ \bibnamefont {Peccei}},\ }\href {https://doi.org/10.1007/978-3-540-73518-2_1} {\bibfield  {journal} {\bibinfo  {journal} {Lect. Notes Phys.}\ }\textbf {\bibinfo {volume} {741}},\ \bibinfo {pages} {3} (\bibinfo {year} {2008})},\ \Eprint {https://arxiv.org/abs/hep-ph/0607268} {arXiv:hep-ph/0607268 [hep-ph]} \BibitemShut {NoStop}%
\bibitem [{\citenamefont {Abbott}\ and\ \citenamefont {Sikivie}(1983)}]{Abbott:1982af}%
  \BibitemOpen
  \bibfield  {author} {\bibinfo {author} {\bibfnamefont {L.}~\bibnamefont {Abbott}}\ and\ \bibinfo {author} {\bibfnamefont {P.}~\bibnamefont {Sikivie}},\ }\href {https://doi.org/10.1016/0370-2693(83)90638-X} {\bibfield  {journal} {\bibinfo  {journal} {Phys. Lett. B}\ }\textbf {\bibinfo {volume} {120}},\ \bibinfo {pages} {133} (\bibinfo {year} {1983})}\BibitemShut {NoStop}%
\bibitem [{\citenamefont {Dine}\ and\ \citenamefont {Fischler}(1983)}]{Dine:1982ah}%
  \BibitemOpen
  \bibfield  {author} {\bibinfo {author} {\bibfnamefont {M.}~\bibnamefont {Dine}}\ and\ \bibinfo {author} {\bibfnamefont {W.}~\bibnamefont {Fischler}},\ }\href {https://doi.org/10.1016/0370-2693(83)90639-1} {\bibfield  {journal} {\bibinfo  {journal} {Phys. Lett. B}\ }\textbf {\bibinfo {volume} {120}},\ \bibinfo {pages} {137} (\bibinfo {year} {1983})}\BibitemShut {NoStop}%
\bibitem [{\citenamefont {Preskill}\ \emph {et~al.}(1983)\citenamefont {Preskill}, \citenamefont {Wise},\ and\ \citenamefont {Wilczek}}]{Preskill:1982cy}%
  \BibitemOpen
  \bibfield  {author} {\bibinfo {author} {\bibfnamefont {J.}~\bibnamefont {Preskill}}, \bibinfo {author} {\bibfnamefont {M.~B.}\ \bibnamefont {Wise}},\ and\ \bibinfo {author} {\bibfnamefont {F.}~\bibnamefont {Wilczek}},\ }\href {https://doi.org/10.1016/0370-2693(83)90637-8} {\bibfield  {journal} {\bibinfo  {journal} {Phys. Lett.}\ }\textbf {\bibinfo {volume} {B120}},\ \bibinfo {pages} {127} (\bibinfo {year} {1983})}\BibitemShut {NoStop}%
\bibitem [{\citenamefont {Svrcek}\ and\ \citenamefont {Witten}(2006)}]{Svrcek:2006yi}%
  \BibitemOpen
  \bibfield  {author} {\bibinfo {author} {\bibfnamefont {P.}~\bibnamefont {Svrcek}}\ and\ \bibinfo {author} {\bibfnamefont {E.}~\bibnamefont {Witten}},\ }\href {https://doi.org/10.1088/1126-6708/2006/06/051} {\bibfield  {journal} {\bibinfo  {journal} {JHEP}\ }\textbf {\bibinfo {volume} {06}},\ \bibinfo {pages} {051}},\ \Eprint {https://arxiv.org/abs/hep-th/0605206} {arXiv:hep-th/0605206} \BibitemShut {NoStop}%
\bibitem [{\citenamefont {Arvanitaki}\ \emph {et~al.}(2010)\citenamefont {Arvanitaki}, \citenamefont {Dimopoulos}, \citenamefont {Dubovsky}, \citenamefont {Kaloper},\ and\ \citenamefont {March-Russell}}]{Arvanitaki:2009fg}%
  \BibitemOpen
  \bibfield  {author} {\bibinfo {author} {\bibfnamefont {A.}~\bibnamefont {Arvanitaki}}, \bibinfo {author} {\bibfnamefont {S.}~\bibnamefont {Dimopoulos}}, \bibinfo {author} {\bibfnamefont {S.}~\bibnamefont {Dubovsky}}, \bibinfo {author} {\bibfnamefont {N.}~\bibnamefont {Kaloper}},\ and\ \bibinfo {author} {\bibfnamefont {J.}~\bibnamefont {March-Russell}},\ }\href {https://doi.org/10.1103/PhysRevD.81.123530} {\bibfield  {journal} {\bibinfo  {journal} {Phys. Rev. D}\ }\textbf {\bibinfo {volume} {81}},\ \bibinfo {pages} {123530} (\bibinfo {year} {2010})},\ \Eprint {https://arxiv.org/abs/0905.4720} {arXiv:0905.4720 [hep-th]} \BibitemShut {NoStop}%
\bibitem [{\citenamefont {Hiramatsu}\ \emph {et~al.}(2013)\citenamefont {Hiramatsu}, \citenamefont {Kawasaki}, \citenamefont {Saikawa},\ and\ \citenamefont {Sekiguchi}}]{Hiramatsu:2012sc}%
  \BibitemOpen
  \bibfield  {author} {\bibinfo {author} {\bibfnamefont {T.}~\bibnamefont {Hiramatsu}}, \bibinfo {author} {\bibfnamefont {M.}~\bibnamefont {Kawasaki}}, \bibinfo {author} {\bibfnamefont {K.}~\bibnamefont {Saikawa}},\ and\ \bibinfo {author} {\bibfnamefont {T.}~\bibnamefont {Sekiguchi}},\ }\href {https://doi.org/10.1088/1475-7516/2013/01/001} {\bibfield  {journal} {\bibinfo  {journal} {JCAP}\ }\textbf {\bibinfo {volume} {01}},\ \bibinfo {pages} {001}},\ \Eprint {https://arxiv.org/abs/1207.3166} {arXiv:1207.3166 [hep-ph]} \BibitemShut {NoStop}%
\bibitem [{\citenamefont {Co}\ \emph {et~al.}(2018)\citenamefont {Co}, \citenamefont {Hall},\ and\ \citenamefont {Harigaya}}]{Co:2017mop}%
  \BibitemOpen
  \bibfield  {author} {\bibinfo {author} {\bibfnamefont {R.~T.}\ \bibnamefont {Co}}, \bibinfo {author} {\bibfnamefont {L.~J.}\ \bibnamefont {Hall}},\ and\ \bibinfo {author} {\bibfnamefont {K.}~\bibnamefont {Harigaya}},\ }\href {https://doi.org/10.1103/PhysRevLett.120.211602} {\bibfield  {journal} {\bibinfo  {journal} {Phys. Rev. Lett.}\ }\textbf {\bibinfo {volume} {120}},\ \bibinfo {pages} {211602} (\bibinfo {year} {2018})},\ \Eprint {https://arxiv.org/abs/1711.10486} {arXiv:1711.10486 [hep-ph]} \BibitemShut {NoStop}%
\bibitem [{\citenamefont {Baratella}\ \emph {et~al.}(2019)\citenamefont {Baratella}, \citenamefont {Pomarol},\ and\ \citenamefont {Rompineve}}]{Baratella:2018pxi}%
  \BibitemOpen
  \bibfield  {author} {\bibinfo {author} {\bibfnamefont {P.}~\bibnamefont {Baratella}}, \bibinfo {author} {\bibfnamefont {A.}~\bibnamefont {Pomarol}},\ and\ \bibinfo {author} {\bibfnamefont {F.}~\bibnamefont {Rompineve}},\ }\href {https://doi.org/10.1007/JHEP03(2019)100} {\bibfield  {journal} {\bibinfo  {journal} {JHEP}\ }\textbf {\bibinfo {volume} {03}},\ \bibinfo {pages} {100}},\ \Eprint {https://arxiv.org/abs/1812.06996} {arXiv:1812.06996 [hep-ph]} \BibitemShut {NoStop}%
\bibitem [{\citenamefont {Co}\ \emph {et~al.}(2020{\natexlab{a}})\citenamefont {Co}, \citenamefont {Hall},\ and\ \citenamefont {Harigaya}}]{Co:2019jts}%
  \BibitemOpen
  \bibfield  {author} {\bibinfo {author} {\bibfnamefont {R.~T.}\ \bibnamefont {Co}}, \bibinfo {author} {\bibfnamefont {L.~J.}\ \bibnamefont {Hall}},\ and\ \bibinfo {author} {\bibfnamefont {K.}~\bibnamefont {Harigaya}},\ }\href {https://doi.org/10.1103/PhysRevLett.124.251802} {\bibfield  {journal} {\bibinfo  {journal} {Phys. Rev. Lett.}\ }\textbf {\bibinfo {volume} {124}},\ \bibinfo {pages} {251802} (\bibinfo {year} {2020}{\natexlab{a}})},\ \Eprint {https://arxiv.org/abs/1910.14152} {arXiv:1910.14152 [hep-ph]} \BibitemShut {NoStop}%
\bibitem [{\citenamefont {Harigaya}\ and\ \citenamefont {Leedom}(2020)}]{Harigaya:2019qnl}%
  \BibitemOpen
  \bibfield  {author} {\bibinfo {author} {\bibfnamefont {K.}~\bibnamefont {Harigaya}}\ and\ \bibinfo {author} {\bibfnamefont {J.~M.}\ \bibnamefont {Leedom}},\ }\href {https://doi.org/10.1007/JHEP06(2020)034} {\bibfield  {journal} {\bibinfo  {journal} {JHEP}\ }\textbf {\bibinfo {volume} {06}},\ \bibinfo {pages} {034}},\ \Eprint {https://arxiv.org/abs/1910.04163} {arXiv:1910.04163 [hep-ph]} \BibitemShut {NoStop}%
\bibitem [{\citenamefont {Co}\ \emph {et~al.}(2020{\natexlab{b}})\citenamefont {Co}, \citenamefont {Hall}, \citenamefont {Harigaya}, \citenamefont {Olive},\ and\ \citenamefont {Verner}}]{Co:2020dya}%
  \BibitemOpen
  \bibfield  {author} {\bibinfo {author} {\bibfnamefont {R.~T.}\ \bibnamefont {Co}}, \bibinfo {author} {\bibfnamefont {L.~J.}\ \bibnamefont {Hall}}, \bibinfo {author} {\bibfnamefont {K.}~\bibnamefont {Harigaya}}, \bibinfo {author} {\bibfnamefont {K.~A.}\ \bibnamefont {Olive}},\ and\ \bibinfo {author} {\bibfnamefont {S.}~\bibnamefont {Verner}},\ }\href {https://doi.org/10.1088/1475-7516/2020/08/036} {\bibfield  {journal} {\bibinfo  {journal} {JCAP}\ }\textbf {\bibinfo {volume} {08}},\ \bibinfo {pages} {036}},\ \Eprint {https://arxiv.org/abs/2004.00629} {arXiv:2004.00629 [hep-ph]} \BibitemShut {NoStop}%
\bibitem [{\citenamefont {Er{\"o}ncel}\ \emph {et~al.}(2022)\citenamefont {Er{\"o}ncel}, \citenamefont {Sato}, \citenamefont {Servant},\ and\ \citenamefont {S{\o}rensen}}]{Eroncel:2022vjg}%
  \BibitemOpen
  \bibfield  {author} {\bibinfo {author} {\bibfnamefont {C.}~\bibnamefont {Er{\"o}ncel}}, \bibinfo {author} {\bibfnamefont {R.}~\bibnamefont {Sato}}, \bibinfo {author} {\bibfnamefont {G.}~\bibnamefont {Servant}},\ and\ \bibinfo {author} {\bibfnamefont {P.}~\bibnamefont {S{\o}rensen}},\ }\href {https://doi.org/10.1088/1475-7516/2022/10/053} {\bibfield  {journal} {\bibinfo  {journal} {JCAP}\ }\textbf {\bibinfo {volume} {10}},\ \bibinfo {pages} {053}},\ \Eprint {https://arxiv.org/abs/2206.14259} {arXiv:2206.14259 [hep-ph]} \BibitemShut {NoStop}%
\bibitem [{\citenamefont {Redi}\ and\ \citenamefont {Tesi}(2023)}]{Redi:2022llj}%
  \BibitemOpen
  \bibfield  {author} {\bibinfo {author} {\bibfnamefont {M.}~\bibnamefont {Redi}}\ and\ \bibinfo {author} {\bibfnamefont {A.}~\bibnamefont {Tesi}},\ }\href {https://doi.org/10.1103/PhysRevD.107.095032} {\bibfield  {journal} {\bibinfo  {journal} {Phys. Rev. D}\ }\textbf {\bibinfo {volume} {107}},\ \bibinfo {pages} {095032} (\bibinfo {year} {2023})},\ \Eprint {https://arxiv.org/abs/2211.06421} {arXiv:2211.06421 [hep-ph]} \BibitemShut {NoStop}%
\bibitem [{\citenamefont {Harigaya}\ and\ \citenamefont {Wang}(2022)}]{Harigaya:2022pjd}%
  \BibitemOpen
  \bibfield  {author} {\bibinfo {author} {\bibfnamefont {K.}~\bibnamefont {Harigaya}}\ and\ \bibinfo {author} {\bibfnamefont {L.-T.}\ \bibnamefont {Wang}},\ }\href@noop {} {\  (\bibinfo {year} {2022})},\ \Eprint {https://arxiv.org/abs/2211.08289} {arXiv:2211.08289 [hep-ph]} \BibitemShut {NoStop}%
\bibitem [{\citenamefont {Er\"oncel}\ \emph {et~al.}(2024)\citenamefont {Er\"oncel}, \citenamefont {Sato}, \citenamefont {Servant},\ and\ \citenamefont {S\o{}rensen}}]{Eroncel:2024rpe}%
  \BibitemOpen
  \bibfield  {author} {\bibinfo {author} {\bibfnamefont {C.}~\bibnamefont {Er\"oncel}}, \bibinfo {author} {\bibfnamefont {R.}~\bibnamefont {Sato}}, \bibinfo {author} {\bibfnamefont {G.}~\bibnamefont {Servant}},\ and\ \bibinfo {author} {\bibfnamefont {P.}~\bibnamefont {S\o{}rensen}},\ }\href@noop {} {\  (\bibinfo {year} {2024})},\ \Eprint {https://arxiv.org/abs/2408.08355} {arXiv:2408.08355 [hep-ph]} \BibitemShut {NoStop}%
\bibitem [{\citenamefont {Er\"oncel}\ \emph {et~al.}(2025)\citenamefont {Er\"oncel}, \citenamefont {Gouttenoire}, \citenamefont {Sato}, \citenamefont {Servant},\ and\ \citenamefont {Simakachorn}}]{Eroncel:2025qlk}%
  \BibitemOpen
  \bibfield  {author} {\bibinfo {author} {\bibfnamefont {C.}~\bibnamefont {Er\"oncel}}, \bibinfo {author} {\bibfnamefont {Y.}~\bibnamefont {Gouttenoire}}, \bibinfo {author} {\bibfnamefont {R.}~\bibnamefont {Sato}}, \bibinfo {author} {\bibfnamefont {G.}~\bibnamefont {Servant}},\ and\ \bibinfo {author} {\bibfnamefont {P.}~\bibnamefont {Simakachorn}},\ }\href@noop {} {\  (\bibinfo {year} {2025})},\ \Eprint {https://arxiv.org/abs/2503.04880} {arXiv:2503.04880 [hep-ph]} \BibitemShut {NoStop}%
\bibitem [{\citenamefont {Bodas}\ \emph {et~al.}(2025)\citenamefont {Bodas}, \citenamefont {Co}, \citenamefont {Ghalsasi}, \citenamefont {Harigaya},\ and\ \citenamefont {Wang}}]{Bodas:2025eca}%
  \BibitemOpen
  \bibfield  {author} {\bibinfo {author} {\bibfnamefont {A.}~\bibnamefont {Bodas}}, \bibinfo {author} {\bibfnamefont {R.~T.}\ \bibnamefont {Co}}, \bibinfo {author} {\bibfnamefont {A.}~\bibnamefont {Ghalsasi}}, \bibinfo {author} {\bibfnamefont {K.}~\bibnamefont {Harigaya}},\ and\ \bibinfo {author} {\bibfnamefont {L.-T.}\ \bibnamefont {Wang}},\ }\href@noop {} {\  (\bibinfo {year} {2025})},\ \Eprint {https://arxiv.org/abs/2503.04888} {arXiv:2503.04888 [hep-ph]} \BibitemShut {NoStop}%
\bibitem [{\citenamefont {Hook}\ \emph {et~al.}(2020)\citenamefont {Hook}, \citenamefont {Marques-Tavares},\ and\ \citenamefont {Tsai}}]{Hook:2019hdk}%
  \BibitemOpen
  \bibfield  {author} {\bibinfo {author} {\bibfnamefont {A.}~\bibnamefont {Hook}}, \bibinfo {author} {\bibfnamefont {G.}~\bibnamefont {Marques-Tavares}},\ and\ \bibinfo {author} {\bibfnamefont {Y.}~\bibnamefont {Tsai}},\ }\href {https://doi.org/10.1103/PhysRevLett.124.211801} {\bibfield  {journal} {\bibinfo  {journal} {Phys. Rev. Lett.}\ }\textbf {\bibinfo {volume} {124}},\ \bibinfo {pages} {211801} (\bibinfo {year} {2020})},\ \Eprint {https://arxiv.org/abs/1912.08817} {arXiv:1912.08817 [hep-ph]} \BibitemShut {NoStop}%
\bibitem [{\citenamefont {Di~Luzio}\ \emph {et~al.}(2021)\citenamefont {Di~Luzio}, \citenamefont {Gavela}, \citenamefont {Quilez},\ and\ \citenamefont {Ringwald}}]{DiLuzio:2021gos}%
  \BibitemOpen
  \bibfield  {author} {\bibinfo {author} {\bibfnamefont {L.}~\bibnamefont {Di~Luzio}}, \bibinfo {author} {\bibfnamefont {B.}~\bibnamefont {Gavela}}, \bibinfo {author} {\bibfnamefont {P.}~\bibnamefont {Quilez}},\ and\ \bibinfo {author} {\bibfnamefont {A.}~\bibnamefont {Ringwald}},\ }\href {https://doi.org/10.1088/1475-7516/2021/10/001} {\bibfield  {journal} {\bibinfo  {journal} {JCAP}\ }\textbf {\bibinfo {volume} {10}},\ \bibinfo {pages} {001}},\ \Eprint {https://arxiv.org/abs/2102.01082} {arXiv:2102.01082 [hep-ph]} \BibitemShut {NoStop}%
\bibitem [{\citenamefont {Jeong}\ \emph {et~al.}(2022)\citenamefont {Jeong}, \citenamefont {Matsukawa}, \citenamefont {Nakagawa},\ and\ \citenamefont {Takahashi}}]{Jeong:2022kdr}%
  \BibitemOpen
  \bibfield  {author} {\bibinfo {author} {\bibfnamefont {K.~S.}\ \bibnamefont {Jeong}}, \bibinfo {author} {\bibfnamefont {K.}~\bibnamefont {Matsukawa}}, \bibinfo {author} {\bibfnamefont {S.}~\bibnamefont {Nakagawa}},\ and\ \bibinfo {author} {\bibfnamefont {F.}~\bibnamefont {Takahashi}},\ }\href {https://doi.org/10.1088/1475-7516/2022/03/026} {\bibfield  {journal} {\bibinfo  {journal} {JCAP}\ }\textbf {\bibinfo {volume} {03}}\bibfield  {number} {\bibinfo  {number} { (03)},\ \bibinfo {pages} {026}},\ }\Eprint {https://arxiv.org/abs/2201.00681} {arXiv:2201.00681 [hep-ph]} \BibitemShut {NoStop}%
\bibitem [{\citenamefont {Papageorgiou}\ \emph {et~al.}(2023)\citenamefont {Papageorgiou}, \citenamefont {Qu\'\i{}lez},\ and\ \citenamefont {Schmitz}}]{Papageorgiou:2022prc}%
  \BibitemOpen
  \bibfield  {author} {\bibinfo {author} {\bibfnamefont {A.}~\bibnamefont {Papageorgiou}}, \bibinfo {author} {\bibfnamefont {P.}~\bibnamefont {Qu\'\i{}lez}},\ and\ \bibinfo {author} {\bibfnamefont {K.}~\bibnamefont {Schmitz}},\ }\href {https://doi.org/10.1007/JHEP01(2023)169} {\bibfield  {journal} {\bibinfo  {journal} {JHEP}\ }\textbf {\bibinfo {volume} {01}},\ \bibinfo {pages} {169}},\ \Eprint {https://arxiv.org/abs/2206.01129} {arXiv:2206.01129 [hep-ph]} \BibitemShut {NoStop}%
\bibitem [{\citenamefont {Di~Luzio}\ and\ \citenamefont {S\o{}rensen}(2024)}]{DiLuzio:2024fyt}%
  \BibitemOpen
  \bibfield  {author} {\bibinfo {author} {\bibfnamefont {L.}~\bibnamefont {Di~Luzio}}\ and\ \bibinfo {author} {\bibfnamefont {P.}~\bibnamefont {S\o{}rensen}},\ }\href {https://doi.org/10.1007/JHEP10(2024)239} {\bibfield  {journal} {\bibinfo  {journal} {JHEP}\ }\textbf {\bibinfo {volume} {10}},\ \bibinfo {pages} {239}},\ \Eprint {https://arxiv.org/abs/2408.04623} {arXiv:2408.04623 [hep-ph]} \BibitemShut {NoStop}%
\bibitem [{\citenamefont {Turner}(1986)}]{Turner:1985si}%
  \BibitemOpen
  \bibfield  {author} {\bibinfo {author} {\bibfnamefont {M.~S.}\ \bibnamefont {Turner}},\ }\href {https://doi.org/10.1103/PhysRevD.33.889} {\bibfield  {journal} {\bibinfo  {journal} {Phys. Rev. D}\ }\textbf {\bibinfo {volume} {33}},\ \bibinfo {pages} {889} (\bibinfo {year} {1986})}\BibitemShut {NoStop}%
\bibitem [{\citenamefont {Strobl}\ and\ \citenamefont {Weiler}(1994)}]{Strobl:1994wk}%
  \BibitemOpen
  \bibfield  {author} {\bibinfo {author} {\bibfnamefont {K.}~\bibnamefont {Strobl}}\ and\ \bibinfo {author} {\bibfnamefont {T.~J.}\ \bibnamefont {Weiler}},\ }\href {https://doi.org/10.1103/PhysRevD.50.7690} {\bibfield  {journal} {\bibinfo  {journal} {Phys. Rev. D}\ }\textbf {\bibinfo {volume} {50}},\ \bibinfo {pages} {7690} (\bibinfo {year} {1994})},\ \Eprint {https://arxiv.org/abs/astro-ph/9405028} {arXiv:astro-ph/9405028} \BibitemShut {NoStop}%
\bibitem [{\citenamefont {Co}\ \emph {et~al.}(2019)\citenamefont {Co}, \citenamefont {Gonzalez},\ and\ \citenamefont {Harigaya}}]{Co:2018mho}%
  \BibitemOpen
  \bibfield  {author} {\bibinfo {author} {\bibfnamefont {R.~T.}\ \bibnamefont {Co}}, \bibinfo {author} {\bibfnamefont {E.}~\bibnamefont {Gonzalez}},\ and\ \bibinfo {author} {\bibfnamefont {K.}~\bibnamefont {Harigaya}},\ }\href {https://doi.org/10.1007/JHEP05(2019)163} {\bibfield  {journal} {\bibinfo  {journal} {JHEP}\ }\textbf {\bibinfo {volume} {05}},\ \bibinfo {pages} {163}},\ \Eprint {https://arxiv.org/abs/1812.11192} {arXiv:1812.11192 [hep-ph]} \BibitemShut {NoStop}%
\bibitem [{\citenamefont {Takahashi}\ and\ \citenamefont {Yin}(2019)}]{Takahashi:2019pqf}%
  \BibitemOpen
  \bibfield  {author} {\bibinfo {author} {\bibfnamefont {F.}~\bibnamefont {Takahashi}}\ and\ \bibinfo {author} {\bibfnamefont {W.}~\bibnamefont {Yin}},\ }\href {https://doi.org/10.1007/JHEP10(2019)120} {\bibfield  {journal} {\bibinfo  {journal} {JHEP}\ }\textbf {\bibinfo {volume} {10}},\ \bibinfo {pages} {120}},\ \Eprint {https://arxiv.org/abs/1908.06071} {arXiv:1908.06071 [hep-ph]} \BibitemShut {NoStop}%
\bibitem [{\citenamefont {Arvanitaki}\ \emph {et~al.}(2020)\citenamefont {Arvanitaki}, \citenamefont {Dimopoulos}, \citenamefont {Galanis}, \citenamefont {Lehner}, \citenamefont {Thompson},\ and\ \citenamefont {Van~Tilburg}}]{Arvanitaki:2019rax}%
  \BibitemOpen
  \bibfield  {author} {\bibinfo {author} {\bibfnamefont {A.}~\bibnamefont {Arvanitaki}}, \bibinfo {author} {\bibfnamefont {S.}~\bibnamefont {Dimopoulos}}, \bibinfo {author} {\bibfnamefont {M.}~\bibnamefont {Galanis}}, \bibinfo {author} {\bibfnamefont {L.}~\bibnamefont {Lehner}}, \bibinfo {author} {\bibfnamefont {J.~O.}\ \bibnamefont {Thompson}},\ and\ \bibinfo {author} {\bibfnamefont {K.}~\bibnamefont {Van~Tilburg}},\ }\href {https://doi.org/10.1103/PhysRevD.101.083014} {\bibfield  {journal} {\bibinfo  {journal} {Phys. Rev. D}\ }\textbf {\bibinfo {volume} {101}},\ \bibinfo {pages} {083014} (\bibinfo {year} {2020})},\ \Eprint {https://arxiv.org/abs/1909.11665} {arXiv:1909.11665 [astro-ph.CO]} \BibitemShut {NoStop}%
\bibitem [{\citenamefont {Huang}\ \emph {et~al.}(2020)\citenamefont {Huang}, \citenamefont {Madden}, \citenamefont {Racco},\ and\ \citenamefont {Reig}}]{Huang:2020etx}%
  \BibitemOpen
  \bibfield  {author} {\bibinfo {author} {\bibfnamefont {J.}~\bibnamefont {Huang}}, \bibinfo {author} {\bibfnamefont {A.}~\bibnamefont {Madden}}, \bibinfo {author} {\bibfnamefont {D.}~\bibnamefont {Racco}},\ and\ \bibinfo {author} {\bibfnamefont {M.}~\bibnamefont {Reig}},\ }\href {https://doi.org/10.1007/JHEP10(2020)143} {\bibfield  {journal} {\bibinfo  {journal} {JHEP}\ }\textbf {\bibinfo {volume} {10}},\ \bibinfo {pages} {143}},\ \Eprint {https://arxiv.org/abs/2006.07379} {arXiv:2006.07379 [hep-ph]} \BibitemShut {NoStop}%
\bibitem [{\citenamefont {Amin}\ and\ \citenamefont {Mirbabayi}(2024)}]{Amin:2022nlh}%
  \BibitemOpen
  \bibfield  {author} {\bibinfo {author} {\bibfnamefont {M.~A.}\ \bibnamefont {Amin}}\ and\ \bibinfo {author} {\bibfnamefont {M.}~\bibnamefont {Mirbabayi}},\ }\href {https://doi.org/10.1103/PhysRevLett.132.221004} {\bibfield  {journal} {\bibinfo  {journal} {Phys. Rev. Lett.}\ }\textbf {\bibinfo {volume} {132}},\ \bibinfo {pages} {221004} (\bibinfo {year} {2024})},\ \Eprint {https://arxiv.org/abs/2211.09775} {arXiv:2211.09775 [hep-ph]} \BibitemShut {NoStop}%
\bibitem [{\citenamefont {Liu}\ \emph {et~al.}(2025{\natexlab{a}})\citenamefont {Liu}, \citenamefont {Hu},\ and\ \citenamefont {Xiao}}]{Liu:2024pjg}%
  \BibitemOpen
  \bibfield  {author} {\bibinfo {author} {\bibfnamefont {R.}~\bibnamefont {Liu}}, \bibinfo {author} {\bibfnamefont {W.}~\bibnamefont {Hu}},\ and\ \bibinfo {author} {\bibfnamefont {H.}~\bibnamefont {Xiao}},\ }\href {https://doi.org/10.1103/PhysRevD.111.023535} {\bibfield  {journal} {\bibinfo  {journal} {Phys. Rev. D}\ }\textbf {\bibinfo {volume} {111}},\ \bibinfo {pages} {023535} (\bibinfo {year} {2025}{\natexlab{a}})},\ \Eprint {https://arxiv.org/abs/2406.12970} {arXiv:2406.12970 [hep-ph]} \BibitemShut {NoStop}%
\bibitem [{\citenamefont {Ling}\ and\ \citenamefont {Amin}(2024)}]{Ling:2024qfv}%
  \BibitemOpen
  \bibfield  {author} {\bibinfo {author} {\bibfnamefont {S.}~\bibnamefont {Ling}}\ and\ \bibinfo {author} {\bibfnamefont {M.~A.}\ \bibnamefont {Amin}},\ }\href@noop {} {\  (\bibinfo {year} {2024})},\ \Eprint {https://arxiv.org/abs/2408.05591} {arXiv:2408.05591 [astro-ph.CO]} \BibitemShut {NoStop}%
\bibitem [{\citenamefont {Long}\ and\ \citenamefont {Venegas}(2024)}]{Long:2024imw}%
  \BibitemOpen
  \bibfield  {author} {\bibinfo {author} {\bibfnamefont {A.~J.}\ \bibnamefont {Long}}\ and\ \bibinfo {author} {\bibfnamefont {M.}~\bibnamefont {Venegas}},\ }\href@noop {} {\  (\bibinfo {year} {2024})},\ \Eprint {https://arxiv.org/abs/2412.14322} {arXiv:2412.14322 [astro-ph.CO]} \BibitemShut {NoStop}%
\bibitem [{\citenamefont {Liu}\ \emph {et~al.}(2025{\natexlab{b}})\citenamefont {Liu}, \citenamefont {Hu},\ and\ \citenamefont {Xiao}}]{Liu:2025lts}%
  \BibitemOpen
  \bibfield  {author} {\bibinfo {author} {\bibfnamefont {R.}~\bibnamefont {Liu}}, \bibinfo {author} {\bibfnamefont {W.}~\bibnamefont {Hu}},\ and\ \bibinfo {author} {\bibfnamefont {H.}~\bibnamefont {Xiao}},\ }\href@noop {} {\  (\bibinfo {year} {2025}{\natexlab{b}})},\ \Eprint {https://arxiv.org/abs/2504.01937} {arXiv:2504.01937 [astro-ph.CO]} \BibitemShut {NoStop}%
\bibitem [{\citenamefont {Ir\v{s}i\v{c}}\ \emph {et~al.}(2017)\citenamefont {Ir\v{s}i\v{c}}, \citenamefont {Viel}, \citenamefont {Haehnelt}, \citenamefont {Bolton},\ and\ \citenamefont {Becker}}]{Irsic:2017yje}%
  \BibitemOpen
  \bibfield  {author} {\bibinfo {author} {\bibfnamefont {V.}~\bibnamefont {Ir\v{s}i\v{c}}}, \bibinfo {author} {\bibfnamefont {M.}~\bibnamefont {Viel}}, \bibinfo {author} {\bibfnamefont {M.~G.}\ \bibnamefont {Haehnelt}}, \bibinfo {author} {\bibfnamefont {J.~S.}\ \bibnamefont {Bolton}},\ and\ \bibinfo {author} {\bibfnamefont {G.~D.}\ \bibnamefont {Becker}},\ }\href {https://doi.org/10.1103/PhysRevLett.119.031302} {\bibfield  {journal} {\bibinfo  {journal} {Phys. Rev. Lett.}\ }\textbf {\bibinfo {volume} {119}},\ \bibinfo {pages} {031302} (\bibinfo {year} {2017})},\ \Eprint {https://arxiv.org/abs/1703.04683} {arXiv:1703.04683 [astro-ph.CO]} \BibitemShut {NoStop}%
\bibitem [{\citenamefont {Rogers}\ and\ \citenamefont {Peiris}(2021)}]{Rogers:2020ltq}%
  \BibitemOpen
  \bibfield  {author} {\bibinfo {author} {\bibfnamefont {K.~K.}\ \bibnamefont {Rogers}}\ and\ \bibinfo {author} {\bibfnamefont {H.~V.}\ \bibnamefont {Peiris}},\ }\href {https://doi.org/10.1103/PhysRevLett.126.071302} {\bibfield  {journal} {\bibinfo  {journal} {Phys. Rev. Lett.}\ }\textbf {\bibinfo {volume} {126}},\ \bibinfo {pages} {071302} (\bibinfo {year} {2021})},\ \Eprint {https://arxiv.org/abs/2007.12705} {arXiv:2007.12705 [astro-ph.CO]} \BibitemShut {NoStop}%
\bibitem [{\citenamefont {Ir\v{s}i\v{c}}\ \emph {et~al.}(2020)\citenamefont {Ir\v{s}i\v{c}}, \citenamefont {Xiao},\ and\ \citenamefont {McQuinn}}]{Irsic:2019iff}%
  \BibitemOpen
  \bibfield  {author} {\bibinfo {author} {\bibfnamefont {V.}~\bibnamefont {Ir\v{s}i\v{c}}}, \bibinfo {author} {\bibfnamefont {H.}~\bibnamefont {Xiao}},\ and\ \bibinfo {author} {\bibfnamefont {M.}~\bibnamefont {McQuinn}},\ }\href {https://doi.org/10.1103/PhysRevD.101.123518} {\bibfield  {journal} {\bibinfo  {journal} {Phys. Rev. D}\ }\textbf {\bibinfo {volume} {101}},\ \bibinfo {pages} {123518} (\bibinfo {year} {2020})},\ \Eprint {https://arxiv.org/abs/1911.11150} {arXiv:1911.11150 [astro-ph.CO]} \BibitemShut {NoStop}%
\bibitem [{\citenamefont {Hiramatsu}\ \emph {et~al.}(2011)\citenamefont {Hiramatsu}, \citenamefont {Kawasaki},\ and\ \citenamefont {Saikawa}}]{Hiramatsu:2010yn}%
  \BibitemOpen
  \bibfield  {author} {\bibinfo {author} {\bibfnamefont {T.}~\bibnamefont {Hiramatsu}}, \bibinfo {author} {\bibfnamefont {M.}~\bibnamefont {Kawasaki}},\ and\ \bibinfo {author} {\bibfnamefont {K.}~\bibnamefont {Saikawa}},\ }\href {https://doi.org/10.1088/1475-7516/2011/08/030} {\bibfield  {journal} {\bibinfo  {journal} {JCAP}\ }\textbf {\bibinfo {volume} {08}},\ \bibinfo {pages} {030}},\ \Eprint {https://arxiv.org/abs/1012.4558} {arXiv:1012.4558 [astro-ph.CO]} \BibitemShut {NoStop}%
\bibitem [{\citenamefont {Kawasaki}\ \emph {et~al.}(2015)\citenamefont {Kawasaki}, \citenamefont {Saikawa},\ and\ \citenamefont {Sekiguchi}}]{Kawasaki:2014sqa}%
  \BibitemOpen
  \bibfield  {author} {\bibinfo {author} {\bibfnamefont {M.}~\bibnamefont {Kawasaki}}, \bibinfo {author} {\bibfnamefont {K.}~\bibnamefont {Saikawa}},\ and\ \bibinfo {author} {\bibfnamefont {T.}~\bibnamefont {Sekiguchi}},\ }\href {https://doi.org/10.1103/PhysRevD.91.065014} {\bibfield  {journal} {\bibinfo  {journal} {Phys. Rev. D}\ }\textbf {\bibinfo {volume} {91}},\ \bibinfo {pages} {065014} (\bibinfo {year} {2015})},\ \Eprint {https://arxiv.org/abs/1412.0789} {arXiv:1412.0789 [hep-ph]} \BibitemShut {NoStop}%
\bibitem [{\citenamefont {Viel}\ \emph {et~al.}(2005)\citenamefont {Viel}, \citenamefont {Lesgourgues}, \citenamefont {Haehnelt}, \citenamefont {Matarrese},\ and\ \citenamefont {Riotto}}]{Viel:2005qj}%
  \BibitemOpen
  \bibfield  {author} {\bibinfo {author} {\bibfnamefont {M.}~\bibnamefont {Viel}}, \bibinfo {author} {\bibfnamefont {J.}~\bibnamefont {Lesgourgues}}, \bibinfo {author} {\bibfnamefont {M.~G.}\ \bibnamefont {Haehnelt}}, \bibinfo {author} {\bibfnamefont {S.}~\bibnamefont {Matarrese}},\ and\ \bibinfo {author} {\bibfnamefont {A.}~\bibnamefont {Riotto}},\ }\href {https://doi.org/10.1103/PhysRevD.71.063534} {\bibfield  {journal} {\bibinfo  {journal} {Phys. Rev. D}\ }\textbf {\bibinfo {volume} {71}},\ \bibinfo {pages} {063534} (\bibinfo {year} {2005})},\ \Eprint {https://arxiv.org/abs/astro-ph/0501562} {arXiv:astro-ph/0501562} \BibitemShut {NoStop}%
\bibitem [{\citenamefont {Villasenor}\ \emph {et~al.}(2023)\citenamefont {Villasenor}, \citenamefont {Robertson}, \citenamefont {Madau},\ and\ \citenamefont {Schneider}}]{Villasenor:2022aiy}%
  \BibitemOpen
  \bibfield  {author} {\bibinfo {author} {\bibfnamefont {B.}~\bibnamefont {Villasenor}}, \bibinfo {author} {\bibfnamefont {B.}~\bibnamefont {Robertson}}, \bibinfo {author} {\bibfnamefont {P.}~\bibnamefont {Madau}},\ and\ \bibinfo {author} {\bibfnamefont {E.}~\bibnamefont {Schneider}},\ }\href {https://doi.org/10.1103/PhysRevD.108.023502} {\bibfield  {journal} {\bibinfo  {journal} {Phys. Rev. D}\ }\textbf {\bibinfo {volume} {108}},\ \bibinfo {pages} {023502} (\bibinfo {year} {2023})},\ \Eprint {https://arxiv.org/abs/2209.14220} {arXiv:2209.14220 [astro-ph.CO]} \BibitemShut {NoStop}%
\bibitem [{\citenamefont {O'Hare}\ \emph {et~al.}(2022)\citenamefont {O'Hare}, \citenamefont {Pierobon}, \citenamefont {Redondo},\ and\ \citenamefont {Wong}}]{OHare:2021zrq}%
  \BibitemOpen
  \bibfield  {author} {\bibinfo {author} {\bibfnamefont {C.~A.~J.}\ \bibnamefont {O'Hare}}, \bibinfo {author} {\bibfnamefont {G.}~\bibnamefont {Pierobon}}, \bibinfo {author} {\bibfnamefont {J.}~\bibnamefont {Redondo}},\ and\ \bibinfo {author} {\bibfnamefont {Y.~Y.~Y.}\ \bibnamefont {Wong}},\ }\href {https://doi.org/10.1103/PhysRevD.105.055025} {\bibfield  {journal} {\bibinfo  {journal} {Phys. Rev. D}\ }\textbf {\bibinfo {volume} {105}},\ \bibinfo {pages} {055025} (\bibinfo {year} {2022})},\ \Eprint {https://arxiv.org/abs/2112.05117} {arXiv:2112.05117 [hep-ph]} \BibitemShut {NoStop}%
\bibitem [{\citenamefont {Amin}\ \emph {et~al.}(2025)\citenamefont {Amin}, \citenamefont {May},\ and\ \citenamefont {Mirbabayi}}]{Amin:2025sla}%
  \BibitemOpen
  \bibfield  {author} {\bibinfo {author} {\bibfnamefont {M.~A.}\ \bibnamefont {Amin}}, \bibinfo {author} {\bibfnamefont {S.}~\bibnamefont {May}},\ and\ \bibinfo {author} {\bibfnamefont {M.}~\bibnamefont {Mirbabayi}},\ }\href@noop {} {\  (\bibinfo {year} {2025})},\ \Eprint {https://arxiv.org/abs/2506.12131} {arXiv:2506.12131 [astro-ph.CO]} \BibitemShut {NoStop}%
\bibitem [{\citenamefont {Husdal}(2016)}]{Husdal:2016haj}%
  \BibitemOpen
  \bibfield  {author} {\bibinfo {author} {\bibfnamefont {L.}~\bibnamefont {Husdal}},\ }\href {https://doi.org/10.3390/galaxies4040078} {\bibfield  {journal} {\bibinfo  {journal} {Galaxies}\ }\textbf {\bibinfo {volume} {4}},\ \bibinfo {pages} {78} (\bibinfo {year} {2016})},\ \Eprint {https://arxiv.org/abs/1609.04979} {arXiv:1609.04979 [astro-ph.CO]} \BibitemShut {NoStop}%
\bibitem [{\citenamefont {Niedermann}\ and\ \citenamefont {Sloth}(2020)}]{Niedermann:2020dwg}%
  \BibitemOpen
  \bibfield  {author} {\bibinfo {author} {\bibfnamefont {F.}~\bibnamefont {Niedermann}}\ and\ \bibinfo {author} {\bibfnamefont {M.~S.}\ \bibnamefont {Sloth}},\ }\href {https://doi.org/10.1103/PhysRevD.102.063527} {\bibfield  {journal} {\bibinfo  {journal} {Phys. Rev. D}\ }\textbf {\bibinfo {volume} {102}},\ \bibinfo {pages} {063527} (\bibinfo {year} {2020})},\ \Eprint {https://arxiv.org/abs/2006.06686} {arXiv:2006.06686 [astro-ph.CO]} \BibitemShut {NoStop}%
\bibitem [{\citenamefont {O'Hare}(2020)}]{AxionLimits}%
  \BibitemOpen
  \bibfield  {author} {\bibinfo {author} {\bibfnamefont {C.}~\bibnamefont {O'Hare}},\ }\href {https://doi.org/10.5281/zenodo.3932430} {\bibinfo {title} {cajohare/axionlimits: Axionlimits}},\ \bibinfo {howpublished} {\url{https://cajohare.github.io/AxionLimits/}} (\bibinfo {year} {2020})\BibitemShut {NoStop}%
\bibitem [{\citenamefont {Lyth}(1992)}]{Lyth:1991ub}%
  \BibitemOpen
  \bibfield  {author} {\bibinfo {author} {\bibfnamefont {D.~H.}\ \bibnamefont {Lyth}},\ }\href {https://doi.org/10.1103/PhysRevD.45.3394} {\bibfield  {journal} {\bibinfo  {journal} {Phys. Rev. D}\ }\textbf {\bibinfo {volume} {45}},\ \bibinfo {pages} {3394} (\bibinfo {year} {1992})}\BibitemShut {NoStop}%
\end{thebibliography}%

\end{document}